\documentclass[usegraphicx,usenatbib]{mn2e}

\pdfoutput=1

\usepackage{float}
\usepackage{amssymb}
\usepackage{color}

\newcommand{\de}{\partial}

\begin{document}
\title[Cancelling out systematic uncertainties]{Cancelling out systematic uncertainties}
\author[Nore\~na et al.]{Jorge Nore\~na$^{1}$\thanks{email:jorge.norena@icc.ub.edu}, Licia Verde$^{1,2}$\thanks{email:liciaverde@icc.ub.edu}, Raul Jimenez$^{1,2}$, Carlos Pe\~na-Garay$^{3}$, Cesar Gomez$^{4}$ \\
$^1$ ICC, University of Barcelona(IEEC-UB), Marti i Franques 1, Barcelona 08028, Spain.\\
$^2$ ICREA\\ 
$^3$ Instituto de F\'isica Corpuscular (CSIC-UVEG), Val\`encia, Spain. \\
$^4$ Physics Department and Instituto de Fisica Teorica UAM/CSIC, 28049 Cantoblanco, Madrid, Spain. \\
}

\maketitle

\begin{abstract}
We present a method to minimize, or even cancel out, the nuisance parameters affecting a measurement. Our approach is   general and can be applied to any experiment or observation. We compare   it with   the bayesian technique   used   to deal with nuisance parameters: marginalization, and show how the method compares and improves by avoiding biases. We illustrate the method with several examples taken from the astrophysics and cosmology world: baryonic acoustic oscillations, cosmic clocks, Supernova Type Ia luminosity distance, neutrino oscillations and dark matter detection. By applying the method we recover some known results but also find some interesting new ones. For baryonic acoustic oscillation (BAO) experiments we show how to combine radial and angular BAO measurements   in order to completely eliminate the  dependence on the sound horizon at radiation drag. In the case of exploiting SN1a as standard candles  we show how the uncertainty in the luminosity distance by a second parameter modeled as a metallicity dependence can be eliminated or greatly reduced. When using cosmic clocks to measure the expansion rate of the universe, we demonstrate how a particular combination of observables nearly removes the metallicity dependence of the galaxy on determining differential ages, thus removing the age-metallicity degeneracy in stellar populations. We hope that these findings will be useful in future surveys to obtain robust constraints on the dark energy equation of state.
\end{abstract}



\section{Introduction}

It is often the case that measurements in an experiment are hampered by systematic uncertainties  or poorly known quantities  that bias or increase the error on the experimental quantity we wish to measure. The usual way to deal with these quantities which we will refer to as  ``nuisance parameters'', is to ``marginalize'' over them, \emph{i.e.} in the bayesian framework, to integrate the likelilhood over the full range allowed by the parameter space of the nuisance parameters. This has the inconvenience that it makes the procedure explicitily dependent on the choice of the prior, it is not guaranteed to be unbiased and it is  often non-optimal. A trivial case where marginalization will fail at providing the optimal solution is the following: imagine we obtain two measurements $x$ and $y$ that depend on two quantities of interest  $\theta_1, \theta_2$  and on a nuisance parameter $n$ in the following form $x=\theta_1/n$ and $y=n \theta_2$. It is obvious that the best way to remove the nuisance parameter $n$ is to form the ratio $r=x y$. This reduces the number of ``data" but removes completely the dependence on the nuisance parameter.  In real life it will not always be the case that one can completely eliminate the nuisance parameter because the functional form of the observables might not allow one to do so, or because there are less observables than nuisance parameters.  It would be of value to have a general prescription that  describes whether  the observables can be combined in such a way as to completely eliminate one or more nuisance parameters or, if exact cancellation is not possible, which combination could minimize their impact on the final measurement.

Today, mitigation of the effect of systematic uncertainties is a crucial issue. This is especially true in cosmology, where, in the era of precision cosmology brought about by the avalanche of data of the last decade, statistical errors   keep shrinking, and   the ultimate error-floor is  often imposed by systematics or nuisance parameters. In the literature, however, there is no generic prescription to address this issue besides marginalization. Still, results obtained by marginalization are often not independent of the systematic effects as we will show in concrete cases. The state of the art in mitigating the effect of systematic uncertainties is represented by some specific examples worked out and applied only to specific cases. For example \cite{Percival} noted that  two quantities  related to the growth of cosmological structures that can be measured from galaxy redshift surveys are   $\beta=f/b$ and $\sigma_{8,gal}=b\sigma_{8,m} $. Here $f=d\ln \delta /d\ln a$ is the logarithmic growth  rate of structures and depends on key cosmological parameters that we want to measure, $b$ is the  galaxy bias and our nuisance parameter as it is poorly known, $\sigma_{8,gal,m}$ denotes the rms fluctuations of a sphere of $8 Mpc/h$ for the galaxies and dark matter respectively and cosmological information through the growth factor  is enclosed in  $ \sigma_{8,m}$.   Similarly to our example above, \cite{Percival}  suggested to use  the  combination $\beta\sigma_{8,g}$ to remove the bias. In reality, when looking in detail at the equations governing redshift space distortions in galaxy redshift  surveys, one finds that the full redshift-space power spectrum depends on different combinations of  $f$, $b$ and $\sigma_{8,m}$ namely $f\sigma_{8,m}$, $fb\sigma_{8,m}^2$, $b\sigma_{8,m}$; thus the data themselves may allow  one to separate  the various parameters. The use of ratios of observables to cancel systematics has been widespread. The list of examples also includes observables with unequal energy ranges so that the ratio of observables cancels the dependences on systematic errors in neutrino detectors \citep{villante} or on the theoretical distribution of dark matter particles \citep{weiner, fox}; there are also specific methods that apply in some particular models of dark energy \citep{trotta}.

Another well-known example is  the standard procedure to minimize the effect of noise bias in Cosmic Microwave Background angular power spectra. The angular power spectrum computed from the auto-correlation of a map produced by a given detector is   given by $C_{\ell}^{meas}=C_{\ell}^{true}+C_{\ell}^{noise}$ where $C_{\ell}^{noise}$ denotes the power spectrum of the  detector noise and is called ``noise bias": a poor knowledge of the noise bias will bias the estimate of the CMB power spectrum. On the other hand the cross power spectrum for maps produced by different, uncorrelated, detectors  does not have noise bias.
  
In any other context a new  prescription must be worked out from scratch using detailed knowledge and intuition of the specific problem.

The aim of this paper is to provide a general algorithm that minimizes, in an unbiased way,  the impact of systematic uncertainties if they can be somewhat parameterized and poorly known quantities on experimental measurements. The method is completely general and can be applied to any experiment where the nuisance parameters can be characterized (at least approximately). After illustrating the approach and deriving the general prescription, we present a suite of applications, of different levels of complexity.  Though our approach is general we will consider for definiteness cases for which the relevant quantities can be modeled as power-laws or linear functions.
   In order to illustrate the method, we concentrate in some problems in astronomy and cosmology and show how the method reduces the impact of systematics.  In particular we address the following problems: 1) measurements of the expansion history of the universe using baryonic acoustic oscillations (BAO). Here we show how different measurement of the BAO scale at different redshifts can be optimally combined to cancel out  uncertainties in the sound horizon scale. 2) Solar neutrinos and the solar metallicity problem. We discuss how to better minimize the impact of
uncertainties due to the solar abundances and due to theoretical inputs like opacities and diffusion. 3) Cosmic clocks  and how to minimize the influence of systematic uncertainties in the metallicity of galaxies while estimating their age.  4) SN Type IA surveys to measure the expansion history of the universe and how to minimize the dependence of the Hubble diagram on a second parameter. We also recover some well-known results such as the cancellation of the noise bias in the measurement of the CMB angular power spectra and the cancellation of astrophysical inputs in dark matter observabations through the use of observables with adjusted energy ranges.

\section{A systematic approach. Analysis and Recipe}
\label{A systematic approach. Analysis and Recipe}
We consider $N$ observables  $O_i$, $i=1,...,N$ that depend on $m$ accurate or interesting quantities $\mu_i$, $i=1,...,m$ and $n$ ``unknown''  or ``biased'' nuisance quantities $\nu_i$, $i=1,...,n$, \emph{i.e.} $O_i (\mu_1, \dots, \mu_m, \nu_1, \dots, \nu_n)$. Our working assumption is that we are ignorant of the mean values of the nuisance parameters $\hat\nu_j$ and their errors $\langle\nu_j\nu_l\rangle$ 
and $\langle\mu_j\nu_l\rangle$. 
Our goal is to find combinations of the observable quantities $f_k(O_{i},...,O_{N})$, $k=1,...,M$ such that they are not affected by our ignorance of the nuisance parameters. 
This requirement implies that a) the variance of $f_k$ is independent of the variance and covariance of the $\nu_i$, guaranteed by the lack of explicit dependence of $f_k$ on $\nu_i$ 
and b) the mean value of $f_k$ should be independent of the mean value of the $\nu_i$, guaranteed by the null total derivative of $f_k$ with respect to the $\nu_i$. These conditions can be written explicitly as a set of first order partial differential equations
\begin{eqnarray}
f_k &=& f_k(O_1,...,O_N) ~;\quad \\ 
\frac{d f_k}{d \nu_i}&=& \sum_{j=1}^N \frac{\de f_k}{\de O_j} \frac{\de O_j}{\de \nu_i} = 0\quad\mathrm{for}\;i=1,...,n\,.
\label{eq:totderiv}
\end{eqnarray}
Note that even if the nuisance parameters have a distribution which is very different from a Gaussian, the condition that $f_k$ have no explicit dependence on $\nu_i$ guarantees that the correlation functions involving an arbitrary number of solutions $f_k$, \emph{i.e.} $\langle f_1^{\alpha_1}...f_M^{\alpha_M} \rangle$ will be independent of any correlation functions involving the nuisance parameters $\nu_i$. If the distribution of the nuisance parameters is assumed to be Gaussian, it is enough to require that the partial derivative of $f_k$ with respect to $\nu_i$ vanishes, $\partial f_k / \partial \nu_i = 0$ for $i = 1,...,n$, in order to guarantee that the correlation functions of $f_k$ are independent of correlations involving the nuisance parameters. Furthermore, in the Gaussian case the correlations between the nuisance parameters and the solutions will vanish $\langle\nu_i f_k\rangle = 0$ due to the vanishing total derivative, but in the general non-Gaussian case they are different from zero $\langle\nu_i f_k\rangle \neq 0$.

A natural interpretation of the recipe defined by eq.~(\ref{eq:totderiv}) is the {\it renormalization group} equation \citep{Wilson}. In this framework, the nuisance {\it unknown} quantities $\nu_{i}$ are interpreted as setting the renormalization group scale. Physical quantities $f_k(\nu_{i},O_{j}(\nu))$ are then defined as those functionals of the observables that are invariant under arbitrary changes of the values of the nuisance i.e as the ones satisfying the generalized renormalization group equation
\begin{equation}
(\frac{\partial}{\partial{\nu_{i}}} + \beta_{i,j}\frac{\partial}{\partial{O_j}})f_k(\nu_{i},O_{j}(\nu)) =0
\end{equation}
with $\beta_{i,j} = \frac{\partial O_{j}}{\partial \nu_{i}}$ the corresponding {\it $\beta$ functions}. In other words, by means of this recipe we identify the physics that is invariant under arbitrary rescalings of the nuisance quantities. This nuisance-independent physics is completely characterized by the {\it nuisance scaling dimensions} $\beta_{i,j}$ of the observables. In the limit case where nuisance can be totally washed out this recipe will unravel the underlaying responsible {\it scale invariance}.

For definitiness, let us show the solutions for the case of $N$ observable quantities $O_i$ which are modeled by power laws of the $n$ nuisance parameters $\nu_j$
\begin{equation}
O_i - \hat O_i = g(\vec\mu)\prod_{j=1}^n(\nu_j - \hat\nu_j)^{a_{ij}}\,,
\label{eq:muPowerLaw}
\end{equation}
where $g$ is some function of $\vec\mu$, a vector containing all other quantities on which the observables depend,  $a_{ij}$ are known exponents and $\hat{\nu}_j$ is the true value of $\nu_j$, which does not necessarily  need to be known. The solution to the system of the 2$n$ first-order partial differential equations is a power law of the observables:
\begin{equation}
f_k = \prod_{i=1}^N(O_i - \hat O_i)^{b^{k}_{i}}\,.
\label{eq:fPowerLaw}
\end{equation}
which leads for each $k$ to a system of $j=1,...,n$ linear algebraic equations for the $i=1,...,N$ unknowns $b^{k}_{i}$. Clearly there are  $k=1,...,M=N - n$ non-trivial independent solutions,
\begin{equation}
\sum_{i=1}^m a_{ij} b^{k}_{i} = 0\,.
\label{eq:solutionPowerLaw}
\end{equation}
After removing the nuisance parameters, we are left with $N-n$ observables. This means that we loose one observable per each nuisance parameter we eliminate and therefore that we can only eliminate $N-1$ nuisance parameters. Compared to the case in which  a prior is imposed on the nuisance parameters and one marginalizes over them, in this approach the resulting  statistical errors on the $\mu_j$ are expected to increase.  However we now obtain a set of observables which is independent of the systematics (and thus we do not rely on any choice of prior for the $\nu_i$). This is generally true and does not depend on our assumption that the relevant quantities can be modeled as power-laws. Note also that any combination of solutions to the differential equations, Eq.~\ref{eq:totderiv}, is also a solution, we simply need to find all independent solutions.

Exactly the same analysis can be repeated for the case in which the observables can be modeled by linear combinations of the nuisance parameters
\begin{equation}
O_i - \hat O_i = g(\vec\mu) + \sum_{j=1}^{n}a_{ij}(\nu_j - \hat\nu_j)\,,
\label{eq:muLinear}
\end{equation}
such that the solution to the system of first order differential equations is given by a linear combination of the observables
\begin{equation}
f_k = \sum_{i = 1}^N b^k_i (O_i - \hat O_i)\,
\end{equation}
and we are once more led to a system of $n$ linear algebraic equation for the $N$ unknowns $b^{k}_i$, with $N - n$ non-trivial solutions,
\begin{equation}
\sum_{i = 1}^m a_{ij}b_i^k = 0\,.
\end{equation}

In Sec.~\ref{Application} we discuss applications of Eq.~\ref{eq:solutionPowerLaw} valid for the 
linear and the power law cases and present the non trivial solutions in cases where we have more observables than nuisance parameters. 

Now we concentrate on problems with fewer observables than nuisance parameters. In this case there is no exact solution for the system of Eq.~\ref{eq:totderiv}. However, in some specific cases that are still of practical interest there may be approximate solutions.  In particular, we aim at problems where observables have 
similar (but not identical) dependences on some of the nuisance parameters (i.e., $a_{ij}\sim a_{il}$ for some $i,j,l$). The main difference from the exact treatment above will be that we cannot impose the full condition 
Eq.\ref{eq:totderiv}, but we should rather minimize the impact of systematics on the new observables. If we shift a given nuisance parameter $\nu_i$ 
by $\Delta\nu_i$ the change in $f_k$ is approximately given by $\mathrm{d} f_k/\mathrm{d} \nu_i\,\Delta\nu_i$. The optimal $f_k$ should 
minimize the square of this quantity with respect to the parameters. We avoid the trivial solution by using Lagrange multipliers, which in the particular 
cases of observables modeled by Eq.~\ref{eq:muPowerLaw} or Eq.~\ref{eq:muLinear} leads to the function
\begin{equation}
\mathcal{L}_k = \sum_{j=1}^{n} \bigg(\frac{\mathrm{d}f_k}{\mathrm{d}\nu_j}\bigg)^2 \Delta\nu_j^2 - \lambda_k \Big(\sum_i (b^{k}_i)^2 - A_k^2\Big)\,,
\label{eq:lagrangian}
\end{equation}
where $\lambda_k$ is a Lagrange multiplier to be solved for, $A_k$ is the norm of the vector $b^{k}_i$, and $\Delta\nu_j$ is the uncertainty on the $j$ nuisance parameter. Let us start by studying the power-law case.
In order to minimize this function, we vary it with respect to $b^{k}_i$ using the expression for $f_k$ as a power law of observables given in Eq.~\ref{eq:fPowerLaw}, which 
leads to the eigenvalue equation
\begin{equation}
\sum_{l=1}^{N} {\cal M}_{il} b^{k}_l = \sum_l\Bigg[f_k^2\sum_{j=1}^{n}\bigg(\frac{\Delta\nu_j}{\nu_j}\bigg)^2 a_{ij}  a_{lj}\Bigg] b^{k}_l = \lambda_k b^{k}_i\,.
\label{eq:eigenvalue}
\end{equation}
where the first equality defines the matrix ${\cal M}$. 
This is our central equation. The Lagrange multiplier $\lambda_k$, which will be the eigenvalues of the matrix ${\cal M}^{k}_{il}$, measures how much the extremal (with respect 
to the nuisance parameters) solution $f_k$ is affected by the nuisance parameters.  This can be seen by multiplying Eq.~\ref{eq:eigenvalue} by $b^k_i$, summing over $i$, and noticing that the left-hand side of the resulting expression is precisely $\sum_j(\mathrm{d} f_k/ \mathrm{d} \nu_j)^2 \Delta \nu_j^2$, thus $(\Delta f_k)^2 = \lambda_k \sum_i (b_i^k)^2$. The eigenvectors correspond to independent combinations of observables, and, if we are interested in minimizing the impact of nuisance parameters, we should choose those eigenvectors corresponding to eigenvalues which are small with respect to typical 
entries of the matrix. In fact, the existence of a small eigenvalue is due to the similar dependence of the observables on a particular nuisance parameter.

We can easily verify that this description is more general and includes the previously discussed cases. If an eigenvalue is zero, there is an independent non-trivial solution which is unaffected by changes of the nuisance parameters.  Indeed it can be easily seen from Eq.\ref{eq:eigenvalue} that there will always be $N - n$ null eigenvalues when $n < N$, and their eigenvectors correspond to the solutions of Eq.~\ref{eq:solutionPowerLaw}. Other eigenvalues are expected to be of the order of typical entries of the matrix $\lambda \sim f^2 a^2$, representing a solution for which the change under variations of the nuisance parameters is large $\Delta f^2 \sim f^2 a^2 \sum_i b_i^2$.  In the case where $n>N$, even if all the eigenvalues are different from zero, small eigenvalues of ${\cal M}^{k}_{il}$ correspond to solutions which change only slightly under shifts of the nuisance parameters and one can take these to be the approximately unbiased combinations. There will be as many independent approximately unbiased quantities as there are small eigenvalues of the matrix. These correspond to the instances where the nuisance parameters appear in Eq.\ref{eq:muPowerLaw} with powers which are almost the same for each $O_i$ ($a_{ij}\sim a_{il}$). 

We can repeat the analysis for the linear case. Minimizing the function defined in Eq.~\ref{eq:lagrangian} and using the expression for $f_k$ as a linear combination of observables we now obtain the following eigenvalue equation
\begin{equation}
\sum_{l=1}^{N} {\cal M}_{il} b^{k}_l = \sum_l\Bigg[\sum_{j=1}^{n}(\Delta\nu_j)^2 a_{ij}  a_{lj}\Bigg] b^{k}_l = \lambda_k b^{k}_i\,,
\label{eq:eigenvaluelin}
\end{equation}
where now the matrix ${\cal M}$ has a different shape, but the same considerations as above apply, {\emph i.e.} we look for small eigenvalues and their corresponding eigenvectors will be solutions that are mildly influenced by the nuisance parameters.

Let us now remark that,
in some problems, the nuisance parameters or the observables are
constrained to satisfy $\ell$ relations, say
$g_i(\nu_1,...,\nu_n,O_1,...,O_N) = 0$, for $i = 1, ..., \ell$. This
implies that they are not independent. The way to treat these cases is
again through the use of Lagrange multipliers, such that when the
nuisance parameters can be cancelled out exactly, the total derivative
equation must be replaced by the following one
\begin{equation}
\frac{\mathrm{d}}{\mathrm{d}\nu_i}\bigg(f^k(O_1,...,O_N) +
\sum_{i=1}^\ell \alpha_i g_i(\nu_1,...,\nu_n,O_1,...,O_N)\bigg) = 0\,,
\end{equation}
where $\alpha_i$ are the Lagrange multipliers.  We will see an
explicit example of this case in \S (\ref{cmbSpectrum}). The way to
do it in the formalism of Eq.~\ref{eq:lagrangian} is that instead of
minimizing the function $\mathcal{L}$ from Eq.~\ref{eq:lagrangian},
one has to include the constraints in the following way
\begin{eqnarray}
\mathcal{L}_k
&=& \sum_{j=1}^{n} \bigg(\frac{\mathrm{d}f_k}{\mathrm{d}\nu_j}\bigg)^2
\Delta\nu_j^2 - \lambda_k \Big(\sum_i (b^{k}_i)^2 - A_k^2\Big)  \nonumber \\
&+&
\sum_{i=1}^\ell \alpha_ig_i(\nu_1,...,\nu_n,O_1,...,O_N)\,,
\end{eqnarray}
where $\alpha_i$ are the Lagrange multipliers used to impose the
conditions $g_i = 0$. One then minimizes $\mathcal{L}_k$ with respect
to $b_i^k$, $\lambda_k$, and $\alpha_i$.

Before ending this section, let us make an important comment on Eq.~\ref{eq:lagrangian}. If one imposes an infinite uncertainty on a nuisance parameter, say $\Delta \nu_1 \rightarrow \infty$ for example, typical eigenvalues of the matrix $\mathcal{M}$ will be infinite. However, if there are combinations of observables that are completely independent of $\nu_1$, some eigenvalues will be finite since for those combinations $\mathrm{d} f_k / \mathrm{d} \nu_1 = 0$, so that one should restrict to the space spanned by eigenvectors corresponding to those finite eigenvalues. In this sense looking for combinations of observables which are completely independent of some nuisance parameter $\nu_1$ is equivalent to assigning an infinite uncertainty to $\nu_1.$

\section{Applications}
\label{Application}

In this section we illustrate several cases where our method is applicable. We start with the study of baryon acoustic oscillations measurements where we assume that  the sound horizon at radiation drag,  which is common to all observations, is the only quantity affected  by systematic  effects. In this case the dependence on the systematic  can be cancelled out  exactly.  A similar problem is that of canceling astrophysical inputs in dark matter observations through the use of observables with adjusted energy ranges so that a nuisance function can be cancelled, as done in \cite{weiner, fox}. Next we consider a case where cancellation of the nuisance parameters cannot be exact: we will then  find approximate unbiased combinations. This is the case for solar neutrino fluxes, where the neutrino flux predictions  in detail depend on many  nuisance parameters but the dependences are similar enough that approximate unbiased combinations can be found.

In the spirit of  increasing  complexity, in the next example we consider a case where there is one systematic common to all observations and on top of that a correction  to this main trend that differs from observation to observation in a well known way. The method will cancel out the dependence on the common systematic and minimaze that of the individual  corrections. This is the case of the cosmic clocks. A similar case is that of SNe 1A where the intrinsic magnitude of the SNe  is unknown: its average value will be constant and in common to all SNe but its actual value depends on an hidden variable (probably metallicity of the host galaxy).

To conclude we consider a case that at first sight may go beyond the scope of our method, where the interesting parameters and the nuisance parameters  are defined through correlation of the observables and not the observables themselves. This is the example of the  instrumental noise in CMB angular power spectrum, where the observables are the temperature fluctuations (or their spherical harmonic transform $a_{\ell m}$) but  it is the angular power spectrum $C_{\ell}$ that carries direct information on  the cosmological  parameters and is affected by  noise bias.

\subsection{Measurement of the expansion history of the Universe with Baryon Acoustic 
Oscillations}
\label{sec:BAOs}

Baryonic acoustic oscilation experiments (BAO) will, eventually, measure the acoustic scale in 
the radial and perperdicular direction in the sky. Let us suppose that one has 
measurements of the BAOs at $N_z$ different redshifts $z_i$. The observables produced by the 
astronomical surveys are
\begin{eqnarray}
\widehat{D}_A(z_i)&=&D_A(z_i)/r_s =\delta\theta/(1+z_i)\, ;\,\, \\
\widehat{H}(z_i)&=&H(z_i) r_s = c \Delta z\,,
\end{eqnarray}
 Here $D_A(z)$ denotes the angular diameter distance  to redshift $z$, $H(z)$ the Hubble parameter at redshift $z$, $\delta\theta$ denotes the  angular measurement of the BAO scale and  $ \Delta z$ the radial measurement; $r_s$ denotes the sound horizon at radiation drag, which is the BAO ``standard ruler".
The sound horizon at radiation drag is determined by CMB observations, for standard cosmologies is affected by a very small error but  its estimate can be  significantly biased for non-standard models (e.g., \cite{eisensteinwhite,debernardis,mangilli}). This consideration motivates us to consider $r_s$ a nuisance parameter.
From eq. \ref{eq:muPowerLaw} we see that we can write $f_k$ as a power law of the observables
\begin{equation}
f_k = \prod_{i=1}^{N_z} \widehat{D}_A(z_i)^{b_i^k}  \prod_{j=N_z+1}^{2N_z} \widehat{H}(z_{j-N_z})^{b_{j}^k}\,,
\end{equation}
so that the values of the exponents $b^k_i$ will be given by solutions to the system of linear 
equations Eq.~\ref{eq:solutionPowerLaw} (or equivalently finding the kernel space of the matrix in Eq.~\ref{eq:eigenvalue}), which we write explicitly for the particular case of two redshifts $z_1, z_2$, 
\begin{equation}
(b_1^k + b_2^k -b_3^k - b_4^k) \frac{f_k}{r_s} = 0\,,
\end{equation}
with solutions
\begin{eqnarray}
f_1 &=&  \widehat{D}_A(z_1) \widehat{H}(z_1)\,,\\
f_2 &=& \widehat{H}(z_2) / \widehat{H}(z_1)\,,\\
f_3 &=& \widehat{D}_A(z_2) / \widehat{D}_A(z_1)\,.
\end{eqnarray}
Any function of these quantities will also be a solution of the differential equation but will 
not contain new information. For example, suppose that one adds $f_4 = \widehat{D}_A(z_2) \widehat{H}(z_2)$ to the 
set of solutions; this will simply be a combination of the solutions we listed, 
indeed $f_4 = f_1 f_2 f_3$. For the case of two redshifts, we obtain only $4 - 1$ independent 
quantities after eliminating our single nuisance parameter $r_s$ from our four observables. 

For measurements of BAOs at $N_z$ different redshifts, the $2N_z - 1$ solutions will be given by all 
the ratios $g = \widehat{D}_A(z_i)/\widehat{D}_A(z_1)$ and $h = \widehat{H}(z_i)/\widehat{H}(z_1)$ plus the combination $f = \widehat{D}_A(z_1)\widehat{H}(z_1)$. 
Note that although it may seem that we obtained that $z_1$ is the pivot scale, this is purely 
arbitrary as one may point out that the same quantities with any $z_i$ as the pivot scale are 
also solutions of the differential equation. This is simply a consequence of the fact that any 
combination of our solutions is also a solution.  Any pivot is equally good since the errors on the parameters will be independent of the choice of pivot under the assumption that they are Gaussian.

\subsubsection{Implications for cosmological parameter estimation}

In this subsection we we use the above findings to forecast errors for the dark energy equation of state 
parameter $w_X(z)$ using the widely used parametrization $w_X = w_0 + w_a (1 - a)$. If the reader is not interested in the details of this analysis, she or he can skip to section \ref{darkMatter}.
In terms of this parametrization, the expansion rate and luminosity distance can be written as
\begin{eqnarray}
H(z) & = & \sqrt{\frac{\Omega_m h^2}{1-\Omega_X}} \\ \nonumber
 & & \!\!\!\!\!\!\!\!\!\!\!\!\!\!\!\!\!\!\!\!\!\!\!\!\!\!\!\!\!\!\!\!\!\!\! 
  \label{eq:H} \times \Bigg[(1+z)^3 (1 - \Omega_X) 
+ \Omega_X(1+z)^{3(1+w_0 + w_a)} e^{-3w_az/(1+z)}\Bigg]^{1/2}\,, \\
D_A(z) &=& \frac{c}{(1+z)}\int_0^z \mathrm{d}z' \frac{1}{H(z')}\,. \label{eq:DA}
\end{eqnarray}
Note that although $\Omega_X$ appears explicitly in these expressions, flatness has been assumed\footnote{Although $\Omega_X + \Omega_m = 1$, the present Hubble rate, parametrized by $h$, is still a free parameter. We follow \cite{seoeisenstein} in using $\Omega_X$ and $\Omega_m h^2$ to parametrize the cosmology dependence.}. 
BAOs surveys measure combinations proportional to $D_A(z)/r_s$ and $H(z) r_s$. Using the Fisher 
matrix formalism, we can forecast errors for these combinations \citep{seoeisenstein} combining measurements at $5$ 
redshifts  bins  of width $\Delta z=0.1$ for $z< 0.5$ for  a survey with specifics similar to those of BOSS-SDSSIII \footnote{http://www.sdss3.org/surveys/boss.php}and $15$ redshifts bins at $z > 0.5$ for a survey with specifics similar to those of  EUCLID \footnote{http://arxiv.org/abs/0912.0914}. We will compare 
three cases: keeping the full set of observations $D_A(z_i)/r_s$, $H(z_i) r_s$ (``full"); keeping only the 
ratios to some pivot scale $\widehat{D}_A(z_i)/\widehat{D}_A(z_p)$, $\widehat{H}(z_i)/\widehat{H}(z_p)$ as is sometimes  advocated (``ratios"); and our ``unbiased'' combinations $\widehat{D}_A(z_i)/\widehat{D}_A(z_p)$, $\widehat{H}(z_i)/\widehat{H}(z_p)$, 
$\widehat{D}_A(z_p)\widehat{H}(z_p)$. For the sake of simplicity, we keep the cosmology fixed and take into account 
only errors on $H_0$ and $r_s$. Using equations \ref{eq:H} and \ref{eq:DA} one can use the 
Fisher matrix for $D_A(z_i)/r_s$ and $H(z_i) r_s$ in order to compute the Fisher matrix (and errors) for 
$w_0, w_a, r_s, H_0$. The results are given in table~\ref{tab:fisherBAO}. In the ``full" case we must assume a prior error on $r_s$ (for example a CMB prior) which we take to have a  30\% error since for some non-standard models the (systematic) error on $r_s$ can be estimated to  be as large as that, see e.g., \cite{mangilli}.  

From the table we see that the statistical errors are always smaller when using the ``full'' set 
as was to be expected (our method is throwing away some information in order to obtain 
quantities which are insensitive to systematic errors, this implies that the statistical errors 
must grow). 
Let us remark that when we consider 
measurements at $20$ different redshifts as specified above, we obtain that the errors on $w_0$ 
and $w_a$ change only by $0.4\%$ and $1\%$ respectively between the ``full'' and ``unbiased'' sets.
Thus, one  is insensitive to systematic errors on $r_s$ at  the price of a modest increase in  errors.
The precise estimate of how much information is 
lost depends on  how big the assumed uncertainty on $ r_s$ is (or how big the systematic errors are), as can be seen on the table for the case in which we assume a 2\% error on $r_s$. Finally, note that when one has measurements at fewer redshifts the difference between the ``ratios'' and ``unbiased'' cases increases dramatically, this is due to the fact that one is throwing away one observable  from a set of just a handful of them, thus loosing comparatively more information.

\begin{table*}
\centering
\begin{tabular}{c|c|c|c}
set & $\sigma_{r_s}/r_s$ & redshifts & $F^{-1}_{ij}(w_0,w_a,\Omega_X, \Omega_m h^2, r_s)$ \\
\hline
full & 0.3 & 20 & 
$\left(
\begin{array}{ccccc}
 0.0579 & -0.215 & -0.00801 & 4.45\times 10^{-7} & 0.981 \\
 - & 0.982 & 0.0371 & -2.87\times 10^{-6} & -6.34 \\
 - & - & 0.00146 & -1.14\times 10^{-7} & -0.252 \\
- & - & - & 1.5\times 10^{-6} & -0.000893 \\
 - & - & - & - & 57.6
\end{array}
\right)$ \\
\hline
unbiased & 0.3 & 20 & 
$\left(
\begin{array}{ccccc}
 0.0584 & -0.218 & -0.00813 & 0. & 0. \\
 - & 1. & 0.0379 & 0. & 0. \\
 - & - & 0.00149 & 0. & 0. \\
 - & - & - & 1.5\times 10^{-6} & 0. \\
 - & - & - & - & 2025.
\end{array}
\right)$ \\
\hline
ratios & 0.3 & 20 & 
$\left(
\begin{array}{ccccc}
 0.0587 & -0.216 & -0.00821 & 0. & 0. \\
 - & 1.01 & 0.0375 & 0. & 0. \\
 - & - & 0.00151 & 0. & 0. \\w
 - & - & - & 1.5\times 10^{-6} & 0. \\
 - & - & - & - & 2025.
\end{array}
\right)$ \\
\hline
full & 0.02 & 20 & 
$\left(
\begin{array}{ccccc}
 0.0435 & -0.121 & -0.0043 & 0.0000136 & 0.133 \\
 - & 0.38 & 0.0132 & -0.0000877 & -0.859 \\
 - & - & 0.000505 & -3.49\times 10^{-6} & -0.0341 \\
 - & - & - & 1.49\times 10^{-6} & -0.000121 \\
 - & - & - & - & 7.8142
\end{array}
\right)$ \\
\hline
unbiased & 0.3 & 10 &
$\left(
\begin{array}{ccccc}
 0.198 & -0.472 & -0.00305 & 0. & 0. \\
 - & 1.63 & 0.036 & 0. & 0. \\
 - & - & 0.00201 & 0. & 0. \\
 - & - & - & 1.5\times 10^{-6} & 0. \\
 - & - & - & - & 2025.
\end{array}
\right)$ \\
\hline
ratios & 0.3 & 10 &
$\left(
\begin{array}{ccccc}
 0.21 & -0.454 & -0.00303 & 0. & 0. \\
 - & 1.66 & 0.0361 & 0. & 0. \\
 - & - & 0.00201 & 0. & 0. \\
 - & - & - & 1.5\times 10^{-6} & 0. \\
 - & - & - & - & 2025.
\end{array}
\right)$ \\
\hline
full & 0.3 & 20 &
$\left(
\begin{array}{ccccc}
 0.171 & -0.738 & -0.0285 & 2.05\times 10^{-6} & 4.53 \\
 - & 3.41 & 0.132 & -0.0000103 & -22.8 \\
 - & - & 0.00519 & -4.07\times 10^{-7} & -0.897 \\
- & - & - & 1.5\times 10^{-6} & -0.000842 \\
 - & - & - & - & 169
\end{array}
\right)$ \\
\hline
unbiased & 0.3 & 20 & 
$\left(
\begin{array}{ccccc}
 0.182 & -0.793 & -0.0307 & 0. & 0. \\
 - & 3.69 & 0.143 & 0. & 0. \\
 - & - & 0.00563 & 0. & 0. \\
 - & - & - & 1.5\times 10^{-6} & 0. \\
 - & - & - & - & 2025.
\end{array}
\right)$
\end{tabular}
\caption{Fisher matrix analysis for the BAOs. We repeat the analysis for the cases in which the Fisher is computed from each the ``full'' set of observables $\{D_A(z_i)/r_s, H(z_i)r_s\}$, the ``ratios'' $\{D_A(z_i)/D_A(z_p), H(z_i)/H(z_p)\}$ including external priors on $r_s$,
and our optimally ``unbiased'' set  $\{D_A(z_i)/D_A(z_p), H(z_i)/H(z_p), D_A(z_p) H(z_p)\}$, as specified in the first column; we have not specified the pivot redshift $z_p$ since any choice gives the same result. We have assumed a typical variance $\sigma_{r_s}/r_s$ of $0.02$ or $0.3$,
 as specified in the second column. 
We show the case in which one has measurements at $20$ redshifts up to redshift $z \approx 2$, and also the case in which one has measurements at $10$ redshifts up to redshift $z \approx 0.8$, as stated in the third column.  
In the fourth column we give the full Fisher matrix with the order of the indexes given by $1 = w_0, 2 = w_a, 3=\Omega_X, 4=\Omega_m h^2, 5 = r_s$, thus for example the element in the first row and second column of a matrix gives an estimate for the correlation between $w_0$ and $w_a$. 
The $2\times 2$ submatrix for $w_0, w_a$ in the unbiased and ratios cases never depends on the assumed variance for $r_s$ or $\Omega_m h^2$ by construction. For most of the cases we assumed a prior on $\Omega_X$ of $\sigma_{\Omega_X}/\Omega_X = 0.01$, except for the las two rows.}
\label{tab:fisherBAO}
\end{table*}

\subsection {Dark Matter}
\label{darkMatter}

We discuss in this section another example of exact cancellation of nuisance parameters, or in general of nuisance functions. In particular, we consider energy dependent observables, which have been used to cancel the dependences on systematic errors in neutrino detectors \citep{villante} or on the theoretical distribution of dark matter particles \citep{weiner,fox}.

Let us consider observables $O_i(E)$ that depend on the integral of some function $f(x)$, that we may not know, within some known range [$x_i$,$y_i$], as
\begin{equation}
O_i(E_{min}<E<E_{max}) = g(\mu_i) F(x_i(E_{min}),y_i(E_{max})) \,.
\end{equation}

The function $f(x)$ in the neutrino case is called the response function and contains the convolution of the neutrino cross section, detector resolution and neutrino flux \citep{villante}. The function $f(x)$ in the case of dark matter detection corresponds to $f(v)/|v|$,  the ratio of the dark matter velocity distribution to the dark matter speed when observing the recoil energy spectrum, and to the integral of that function times the dark matter speed when observing total rates \citep{weiner,fox}.

For simplicity, we concentrate on the case of two observables $O_1$. and $O_2$. Let's assume that the observables have a energy range overlap such that we can find energies $E_a, E_b, E_c$ and $E_d$ such that $x_1(E_a)=x_2(E_c)=x$ and $y_1(E_b)=y_2(E_d)=y$.

Our observables can be rewritten as, 
\begin{equation}
O_1(E_{a}<E<E_{b}) = g(\mu_1) F(x,y) 
\end{equation}
\begin{equation}
O_2(E_{c}<E<E_{d}) = g(\mu_2) F(x,y). 
\end{equation}

Following the lines discussed in the previous section, the dependence on the nuisance function $f(x,y)$ is removed by the  combination $O_1(E_{a}<E<E_{b})/O_2(E_{c}<E<E_{d})$, as advocated in the above references.

\subsection {Solar Neutrinos}

The Standard Solar Model (SSM) depends on $\sim$ 20 input parameters, including the solar age and luminosity, the opacity, the rate of diffusion, the zero-age abundances of key elements 
(He, C, N, O, Ne, Mg, Si, S, Ar, Fe), and the nuclear cross sections (S-factors) for the pp chain and CN cycle reactions.  The observable quantities in this case are the solar neutrino fluxes\footnote{The index $i$ here labels the nuclear reaction that produces the neutrinos.} $\phi_i$, in particular the $^7$Be or $^8$B neutrino fluxes, which have been independently  measured by the solar neutrino detectors. The observed neutrino fluxes depend on  elements abundances,  solar structure parameters and on the S-factors, but the quantities of interest are actually  the S-factors and the rest are poorly determined -nuisance- parameters.

To be more specific,  the multi-dimensional set of variations in SSM input parameters 
$\{ \Delta \beta_j \}$  from the SSM best values
 $\{ \beta_j^\mathrm{SSM} \}$ often collapses to a one-dimensional dependence of the neutrino fluxes  on the solar core temperature ($T_c$), where $T_c$ is an implicit function of the variations $\{ \Delta \beta_j \}$ \citep{bahcall}. The correlations between $\phi_i$ and $T_c$ are strong but not exact.
 
 The sensitivity to parameter variations can 
be expressed in terms of the logarithmic partial derivatives $\alpha(i,j)$ evaluated for each neutrino flux  $\phi_i$ and
and each SSM input parameter $\beta_j$,
\begin{equation}
\alpha(i,j) \equiv {\partial \ln{\left[ \phi_i/\phi_i^\mathrm{SSM} \right]} \over \partial \ln{\left[ \beta_j /
\beta_j^\mathrm{SSM}\right]}}
\end{equation}
where $\phi_i^\mathrm{SSM}$ and $\beta_j^\mathrm{SSM}$ denote the SSM best values.
This information, in combination with the assigned uncertainties in the $\beta_j$, then provides
an estimate of the uncertainty in the SSM prediction of $\phi_i$.   Here we employ the 
logarithmic partial derivatives of \cite{PS2008,SHP}, corresponding to the higher metal composition \citep{GS}
SSM,
\begin{eqnarray}
&&f(^{7}\mathrm{Be}) = {\phi(^{7}\mathrm{Be}) \over \phi(^{7}\mathrm{Be})^\mathrm{SSM}} =  \\ 
&& \left[ L_\odot^{3.558} O^{1.267} A^{0.786} D^{0.136} \right]   \nonumber \\
&\times&  \left[ \mathrm{S}_{11}^{-1.07} ~\mathrm{S}_{33}^{-0.441} ~\mathrm{S}_{34}^{0.878}
~\mathrm{S}_{17}^{0.0}~ \mathrm{S}_{e7}^{0.0} ~\mathrm{S}_{114}^{-0.001} \right]   \nonumber \\
&\times& \left[ x_C^{0.004} x_N^{0.002} x_\mathrm{O}^{0.053} x_\mathrm{Ne}^{0.044} x_\mathrm{Mg}^{0.057}
x_\mathrm{Si}^{0.116} x_\mathrm{S}^{0.083} x_\mathrm{Ar}^{0.014} x_\mathrm{Fe}^{0.217} \right] \nonumber
\label{eq:7Be}
\end{eqnarray}
\begin{eqnarray}
&&f(^{8}\mathrm{B}) ={\phi(^{8}\mathrm{B}) \over \phi(^{8}\mathrm{B})^\mathrm{SSM}} =  \\
&&\left[ L_\odot^{7.130} O^{2.702} A^{1.380} D^{0.280} \right]   \nonumber \\
&\times&  \left[ \mathrm{S}_{11}^{-2.73} ~\mathrm{S}_{33}^{-0.427} ~\mathrm{S}_{34}^{0.846}
~\mathrm{S}_{17}^{1.0}~ \mathrm{S}_{e7}^{-1.0} ~\mathrm{S}_{114}^{0.005} \right]   \nonumber \\
&\times& \left[ x_C^{0.025} x_N^{0.007} x_\mathrm{O}^{0.111} x_\mathrm{Ne}^{0.083} x_\mathrm{Mg}^{0.106}
x_\mathrm{Si}^{0.211} x_\mathrm{S}^{0.151} x_\mathrm{Ar}^{0.027} x_\mathrm{Fe}^{0.510} \right] \nonumber
\label{eq:8B}
\end{eqnarray}
where each parameter on the left-hand side represents a $\beta_j/\beta_j^\mathrm{SSM}$.
The luminosity, opacity, solar age, and the diffusion parameters are denoted by
$L_\odot$, $O$, $A$, and $D$, while S and $x$ denote S-factor and abundance ratios.
The errors assigned to the solar model inputs are (0.4, 2.5, 0.44, 15.0, 0.9, 4.3, 5.1, 7.5, 2.0, 7.2, 29.7, 32.0, 38.7, 53.9, 11.5, 11.5, 9.2, 49.6, 11.5)\%\citep{SHP} for all the quantities  $L_\odot$, $O$, $A$,  $D$, $S_j$ $x_q$ in the order as they appear in Eqs. \ref{eq:7Be} -- \ref{eq:8B}.

A new re-evaluation of the solar composition \citep{AGS}, which leads to significantly smaller abundances than 
previously estimated, has led to the lack of matching of helioseismological data and SSM predictions. It is therefore relevant to test observables that are weakly dependent both on solar composition and more theoretical inputs like diffusion and opacity.
In our analysis, we consider $L_\odot$, $O$, $A$, and $D$ and $x$ as nuisance parameters.
The matrix $M^{k}_{il}$ of Eq.~\ref{eq:eigenvalue} for the two combination of  observables $f(^{7}\mathrm{Be})$  and 
$f(^{8}\mathrm{B})$ is 
\[ \left( \begin{array}{cc}
35.69 & 74.42\\
74.42 & 156.34 \end{array} \right)\] 
whose smaller eigenvalue (0.22) has the corresponding eigenvector is proportional to (2.098, 1). Therefore the combination of observables 
 \begin{equation}
\frac{f(^{7}\mathrm{Be})^{2.10}}{f(^{8}\mathrm{B})} =  \left[ \mathrm{S}_{11}^{0.485} ~\mathrm{S}_{33}^{-0.498} ~\mathrm{S}_{34}^{0.996}~\mathrm{S}_{17}^{-1.0}~ \mathrm{S}_{e7}^{1.0} ~\mathrm{S}_{114}^{-0.007} \right]
 \end{equation}
 minimizes the impact of the errors due to composition, opacity and diffusion inputs (0.47\% error) and therefore
 optimizes the determination of nuclear cross sections at solar core temperatures. 

The formalism used here is very quick in testing the robustness of the method to changes in the assumed errors on the nuisance parameters. We have checked that the optimal observable does not significantly change by modifying the errors on some of the nuisance parameters. For example, if we double the errors on the abundance ratios, the power in $f(^{7}\mathrm{Be})$ changes from 2.098 to 2.099 and the error from 0.47\% to 0.92\%, while if we double the errors in theoretical inputs like opacity and diffusion, the power in $f(^{7}\mathrm{Be})$ changes from 2.098 to 2.105 and the error from 0.47\% to 0.48\%. 

\subsection{The expansion history of the universe from Cosmic Clocks}
\label{sec:metallicities}

A direct method to determine the expansion history of the universe is to use cosmic clocks 
\citep{cosmicclock}. The determination of the Hubble parameter is done by estimating the 
differential age of ``clocks", namely passively evolving galaxies. In order to estimate the 
differential age one approach is to  use the spectral feature around the $4000$\AA\, break \citep{moresco}, although it is desirable to exploit the whole galaxy spectrum \citep{jimenez,simon,stern}. This feature, 
called $D4000$, depends both on the age and also on metallicity of the stellar population.  Although the dependence of $D4000$ on the metallicity is weaker than the dependence on age, the metallicity is usually poorly known, and thus acts as a nuisance parameter in obtaining the differential age. To a good approximation the $D4000$ feature can be written as
\begin{equation}
D4000 \propto {\rm age}^{\alpha} Z^{\beta},
\end{equation}  
where $Z$ is the metallicity of the ``clock" and $\alpha, \beta$ are the exponents of the 
corresponding power-laws. For spectra which have a poor signal to noise there will be a large systematic error on $Z$ due to the fact that spectral lines will be harder to identify. This systematic can be modeled to be a signal-to-noise dependent shift $Z^{measured}  = \Delta(\gamma) Z^{true}$ where $\gamma$ is the signal to noise ratio. Of course there might be other systematics affecting what we have called $Z^{true}$, but this is beyond the scope of this paper.
The biggest error is expected to be the systematic shift  in the estimated metallicity $\Delta(\gamma)$, which we model as a power-law
$\Delta(\gamma) = \Theta \gamma^b$,
where we have parametrized our ignorance with $\Theta$ and $b$ \citep{moresco}. We can rewrite this as
\begin{equation}
\Delta(\gamma) = \Theta B^{\log(\gamma_i)}\,,
\label{eq:clocks0}
\end{equation}
 where $B = e^b$. We then describe the ``cosmic clocks" in the
following way:
\begin{equation}
d_i = {\rm age}_i^\alpha {\bar Z}^ \beta (\Theta B^{\log(\gamma_i)})^ \beta\qquad \mathrm{for\;each\;} i\,,
\label{eq:clocks}
\end{equation}
where $d_i$ is the label corresponding to a given absorption line (the D4000 feature in this example), the index $i$ runs over galaxies that have approximately the same metallicity, $\bar{Z}$ is some central value of 
the metalicity, $\gamma_i$ is the signal to noise ratio of each measurement, and $ \beta$ is some known power--thus if the original sample shows a wide range of metallicity it should be split in metallicity bins before it can be described by  Eq.~\ref{eq:clocks}. We now wish to find quantities which are independent of the nuisance parameters: $\Theta$ and $B$. 
In this case  the total derivative 
condition Eq.~\ref{eq:totderiv} becomes
\begin{equation}
\frac{\mathrm{d}f_k}{\mathrm{d} \Theta}=\sum_{i=1}^N \frac{\partial f_k}{\partial d_i} \beta \frac{d_i}{\Theta} = 0\,,
\label{eq:clocksTotDeriv1}
\end{equation}
\begin{equation}
\frac{\mathrm{d}f_k}{\mathrm{d}B} = \sum_{i=1}^N \frac{\partial f_k}{\partial d_i} \beta\log(\gamma_i)\frac{d_i}{B} = 0\,.
\label{eq:clocksTotDeriv2}
\end{equation}
which, if $f_k$ is a power-law $f_k = \prod_i d_i^{b^k_i}$, is simply
\begin{equation}
\sum_{i=1}^N b^k_i = 0\,,
\end{equation}
\begin{equation}
\sum_{i=1}^N b^k_i \log(\gamma_i)= 0\,.
\end{equation}

We have done this explicitly for the example data in table \ref{tab:clocks} for 10 galaxies. This  sample has characteristics not too dissimilar from those of the next $H(z)$ release (Moresco et al. in prep.). The resulting combinations are the eight that satisfy
\begin{equation}
b^k_{9} = \frac{1}{5} (-30 b_1-30 b_2-30 b_3-2 b_4-3 b_5-5 b_6-15 b_7-23 b_8) \label{eq:clocksSol1}
\end{equation}
\begin{equation}
b^k_{10} = 5 b_1+5 b_2 + 5 b_3 -\frac{3 b_4}{5}-\frac{2 b_5}{5}+2 b_7+\frac{18 b_8}{5} \label{eq:clocksSol2}
\end{equation}
%
\begin{table}
\centering
\begin{tabular}{c|c|c|c}
d4000 &    $ \sigma_d$ &         $z$ &    $\gamma$ \\
\hline\hline   
2.00 &          0.03  &        0.15 &  100\\
1.95 &          0.03  &        0.2 &   100\\
1.90 &          0.03  &        0.25 &  100\\
1.81 &           0.05 &          0.4 &      72\\
1.82 &          0.05  &         0.5 &      73\\
1.76 &          0.05  &         0.6 &     75\\
1.83 &          0.03  &         0.7 &    85\\
1.70 &          0.025 &        0.9 &    93\\
1.60 &          0.04 &          1.05 &  75\\
1.50 &          0.05 &          1.28 &  70\\
\hline
\end{tabular}
\caption{Measurements of the spectral feature around the $4000$\AA\, break for different galaxies. These are examples of the types of data that one should obtain, and the results from actual measruements are expected to be similarly distributed. We show the measured value of the D4000 feature in the first column, the error on each measurement on the second column as given by its dispersion, the redshift $z$ of the galaxy in the third column, and the signal to noise of the spectrum in the fourth column.}
\label{tab:clocks}
\end{table}
We explicitly give a set of 8 independent solutions in table \ref{tab:clocksSol}. One may then use these combinations to produce robust measurements of the redshift as a function of cosmic time and apply it for example to the case of dark energy. Of course this cancelation of the systematics depends on the way we model them, but there  is no method to treat completely unknown systematic errors. This method can be generalised to the more general case where the full spectrum is used \citep{jimenez,simon,stern}, although in this case the effect of the metallicity as a systematic will be smaller.
\begin{table}
\centering
\begin{tabular}{c|c|c|c|c|c|c|c|c|c}
$b_1$ & $b_2$ & $b_3$ & $b_4$ & $b_5$ & $b_6$ & $b_7$ & $b_8$ & $b_9$ & $b_{10}$ 
\\ \hline\hline
1 & 0 & 0 & 0 & 0 & 0 & 0 & 0 & -6 & 5 \\ 
0 & 1 & 0 & 0 & 0 & 0 & 0 & 0 & -6 & 5 \\ 
0 & 0 & 1 & 0 & 0 & 0 & 0 & 0 & -6 & 5 \\
0 & 0 & 0 & 1 & 0 & 0 & 0 & 0 & -2/5 & -3/5 \\ 
0 & 0 & 0 & 0 & 1 & 0 & 0 & 0 & -3/5 & -2/5 \\ 
0 & 0 & 0 & 0 & 0 & 1 & 0 & 0 & -1 & 0 \\ 
0 & 0 & 0 & 0 & 0 & 0 & 1 & 0 & -3 & 2 \\ 
0 & 0 & 0 & 0 & 0 & 0 & 0 & 1 & -23/5 & 18/5\\ \hline
\end{tabular}
\caption{Explicit solutions of equations \ref{eq:clocksTotDeriv1} and \ref{eq:clocksTotDeriv2} satisfying conditions \ref{eq:clocksSol1} and \ref{eq:clocksSol2}, for $f_k = \prod_i d_i^{b_i^k}$, using the measurements given in table \ref{tab:clocks}. $k$ is the row index, and the column index $i$ is ordered as in table \ref{tab:clocks}, \emph{i.e.} $i = 1$ correspond to the first row in that table, $i = 2$ to the second, and so on.}
\label{tab:clocksSol}
\end{table}
 
 \subsection{Hubble diagram from SN1a:  Removing second parameter dependence}

The above example was motivated by the cosmic clocks problem, but to show that the solution found is general enough to be applied in different contexts, we consider a different problem in this section. Supernovae type 1A (SN1a) are considered to be standard candles, so that, for a  sample of supernovae spanning a wide redshift interval their apparent magnitude can be used to infer the luminosity distance  as a function of redshift and thus constrain cosmological parameters.

In reality, SN1a are standardizable candles: an empirical algorithm \citep{Phillips} relates peak luminosity 
and stretch of the light curve. 
Using this procedure one can exploit SN1a as if 
they were standard candles; the scatter around the luminosity distance relation gets dramatically reduced,  and thus obtain a luminosity distance $d_L$ using the distance modulus $\mu$:
\begin{equation}
\mu= m-M=25+5\log d_L
\end{equation}
where $d_L$  encloses the dependence on cosmological parameters, $m$ denotes the SN apparent magnitude and $M$ the absolute magnitude. 
The  distance modulus is related to the peak supernova magnitude in $r-$band by
\begin{equation}
m_{\rm peak} = M_{\rm stand} - \alpha (s-1) + \mu (z) + K(s,z) + A_r
\end{equation} 
where $M_{\rm stand}$ is the standardized peak $B$ absolute magnitude, $\alpha$ is the Phillips (1993) parameter relating stretch and peak luminosity, $s$ is a typical stretch, $\mu (z)$ is the distance modulus, $A_r$ is the $r$ band Milky Way extinction along the supernova's line of sight and $K(s,z)$ is an approximate K-correction to the r band at z = 0. Of course, one could use any optical band besides $r$.
All these corrections have a residual scatter and there is evidence that some of the residual scatter could be due to some ``second parameter" (e.g., \cite{Aubourg,Sullivan,Brandt}) such as metallicity of the host galaxy. There are thus two sources of errors: a) the fact that, after all the corrections have been applied the average  absolute magnitude $\bar{M}$, although common to all supernovae in the sample, is unknown and b) that there is a residual scatter around $M$ which could, in principle, not be purely statistical but depend on extra parameters.
 
Thus we can recast the above problem as
\begin{equation}
\hat m_i=\bar{M}+\Delta M_i+\mu_i
\end{equation}
where $i=1,N$ runs over the SN in the sample, $\Delta M_i$ include all residual systematic errors, $\hat m_i$ the (k-corrected, extinction corrected, stretch-corrected) observable quantities and $\mu_i$ enclose the cosmological dependence and is the quantity of interest.
As these are logarithmic quantities, the reader will inmediately realize that this problem is similar to the cosmic clock one  where $\log \Delta=\Delta M_i$, $Z=\log \bar{M}$ and $\log \alpha =\mu_i$. One can completely cancel the dependence on $\bar{M}$ by taking linear combinations $f_k = \sum_i b^k_i m_i$ that satisfy 
\begin{equation}
\sum_i b^k_i = 0\,,
\end{equation}
similarly to the BAO case. One thus obtains combinations that are robust under changes in the estimated value of $\bar M$.

We can do even better and attempt to cancel the remaining systematics $\Delta M_i$. There are indications that the absolute magnitude depends also on the metallicity of the supernova (or host galaxy) and that this is the leading contribution to  $\Delta M_i$, so we will assume that $\Delta M_i$ depends on the host galaxy metallicity $Z$ and model $\Delta M(Z)$ as a polynomial\footnote{The degree of the polynomial can be tuned such that it provides a good description to the actual data.} of some power $\ell$.
\begin{equation}
\Delta M(Z) = \Theta_1 (Z - Z_0) + \Theta_2 (Z - Z_0)^2 + ... + \Theta_\ell (Z - Z_0)^\ell\,,
\end{equation}
where $\Theta_i$ for $i = 1, ..., \ell$ and $Z_0$ are unknown parameters. This can also be thought of as a Taylor expansion of the dependence of $\Delta M$ on $Z$. This polynomial can be explicitly expanded and rewritten in the following form
\begin{equation}
\Delta M(Z) = \tilde\Theta_0 + \tilde\Theta_1 Z + \tilde\Theta_2 Z^2 + ... + \tilde\Theta_\ell Z^\ell\,,
\end{equation}
where $\tilde\Theta_i$ are combinations of $\Theta_j$ and $Z_0$.  There is an ongoing effort to measure the metalicities, so that here we take them to be part of our space of observables. Therefore, our nuisance parameters will be $\bar M$ and $\tilde\Theta_i$ for $i = 0, 1, ..., \ell$. We find that the total derivative conditions can then be written as
\begin{eqnarray}
\frac{\mathrm{d} f_k}{\mathrm{d} \tilde\Theta_0} \propto \frac{\mathrm{d} f_k}{\mathrm{d} \bar M} \propto \sum_{i=1}^N\frac{\partial f_k}{\partial \hat m_i} &=& 0 \,,\\
\frac{\mathrm{d} f_k}{\mathrm{d} \tilde\Theta_1} = \sum_{i=1}^N\frac{\partial f_k}{\partial \hat m_i} Z_i &=& 0 \,,\\
\frac{\mathrm{d} f_k}{\mathrm{d} \tilde\Theta_2} = \sum_{i=1}^N\frac{\partial f_k}{\partial \hat m_i} Z_i^2 &=& 0 \,,\\
&\vdots& \nonumber\\
\frac{\mathrm{d} f_k}{\mathrm{d} \tilde\Theta_\ell} = \sum_{i=1}^N\frac{\partial f_k}{\partial \hat m_i} Z_i^\ell &=& 0 \,.
\end{eqnarray}
 The solution to the set of differential equations is given by a linear combination $f_k = \sum_i b^k_i \hat m_i$, where the coefficients $b^k_i$ are the $N - \ell - 1$ that satisfy the following conditions
 \begin{eqnarray}
 \sum_{i=1}^N b_i^k &=& 0\,,\\
 \sum_{i=1}^N b_i^k Z_i&=& 0\,,\\
 \sum_{i=1}^N b_i^k Z_i^2&=& 0\,,\\
 &\vdots&\nonumber\\
 \sum_{i=1}^N b_i^k Z_i^\ell&=& 0\,.
 \end{eqnarray}
 We expect the results obtained here to be robust also under changes in the $\Theta_\ell$, $\bar M$, and $Z_0$.
 
 Now let us compare  the solution provided by this method  with the standard  approach.
When analyzing a SN sample it is customary to marginalize over $\bar{M}$ as follows: for each choice of cosmological parameters ($\alpha$) $\mu(z_i)$ is computed and used to extract $\bar{M}(\alpha)$;  $\Delta M_i$ then encloses the statistical error. One can see that this procedure is equivalent to marginalize over $H_0$. In fact one can rewrite the apparent magnitude as:
\begin{equation}
m_i=M-5\log H_0 +25+5\log (H_0d_L)_i\,.
\end{equation}

This procedure cannot be unbiased if the set of cosmological models/parameters scanned does not include the true underlying model. The solution provided by the present method instead cancels out the dependence on $\bar{M}(\alpha)$ providing an unbiased answer, and can also mitigate the impact of remaining systematics (such as the influence of the metalicity of on the absolute magnitude). Of course this cancelation of the remaining systematics depends on the way we model them, but there is now way of treating completely unknown systematic uncertainties.

\subsection{CMB angular power spectrum.}
\label{cmbSpectrum}
The angular power spectrum computed from the auto-correlation of a map produced by a given detector is   given by $C_{\ell}^{meas}=C_{\ell}^{true}+C^{noise}_{\ell}$ where $C^{noise}_{\ell}$ denotes the power spectrum of the  detector noise and is called ``noise bias": a poor knowledge of the noise bias will bias the estimate of the CMB power spectrum. On the other hand the cross power spectrum for maps produced by different, uncorrelated, detectors  does not have noise bias. In this subsection we wish to reproduce this well-known fact in the light of our approach. This is a simple example of a problem which is neither linear nor a power-law, and can thus show how to apply our approach in more general cases.

Let us model this case as having two detectors that give two different measurements of the temperature perturbations of the CMB with uncorrelated noise
\begin{equation}
\begin{array}{rcl}
a^{(1)}_{\ell m} &=& a_{\ell m} + N^{(1)}_{\ell m}\,,\\
a^{(2)}_{\ell m} &=& a_{\ell m} + N^{(2)}_{\ell m}\,,
\end{array}
\end{equation}
where $a^{(1,2)}_{\ell m}$ are the temperature anisotropies decomposed in spherical harmonics measured by each detector, $a_{\ell m}$ are the ``true'' temperature anisotropies, and $N^{(1,2)}_{\ell m}$ is the noise of each detector. We wish to apply the conditions given in Eq.~\ref{eq:totderiv} to this case. Since there is a different contribution of the noise per each $m$, it would appear that the problem has no solution as there are as many nuisance parameters as observables. However not all of them are independent since they are uncorrelated among themselves and with the ``true'' temperature anisotropies, and must thus satisfy the following constraints
\begin{eqnarray}
\sum_m N^{(1)}_{\ell m} N^{(2)}_{\ell' m} &=& 0\,,\\
\sum_m N^{(1)}_{\ell m} a_{\ell' m} &=& 0\,, \\
\sum_m N^{(2)}_{\ell m} a_{\ell' m} &=& 0\,.
\end{eqnarray}
One way to approach the problem is to use all these constraints in order to remain only with independent quantities. Though this gives the desired result, it is rather involved algebraically. We will instead use Lagrange multipliers and impose that the total derivative of the following function with respect to each $N^{(1,2)}_{\ell m}$ be zero while keeping all of them independent
\begin{equation}
f_k + \lambda_1^k \sum_m N^{(1)}_{\ell m} N^{(2)}_{\ell m} + \lambda_2^k \sum_m N^{(1)}_{\ell m} a_{\ell m} + \lambda_3^k \sum_m N^{(2)}_{\ell m} a_{\ell m}\,,
\end{equation}
which gives the following two equations
\begin{eqnarray}
\frac{\partial f_k}{\partial a_{\ell m}^{(1)}}  + \lambda_1^k a_{\ell m} + \lambda_2^k N^{(2)}_{\ell m} &=& 0\,,\\
\frac{\partial f_k}{\partial a_{\ell m}^{(2)}}  + \lambda_1^k a_{\ell m} + \lambda_3^k N^{(1)}_{\ell m} &=& 0\,.
\end{eqnarray}
Imposing that the $f_k$ depend explicitly only on $a^{(1,2)}_{\ell m}$ implies $\lambda^k_1 = \lambda^k_2 = \lambda^k_3$, and the differential equations can be rewritten as
\begin{eqnarray}
\frac{\partial f_k}{\partial a_{\ell m}^{(1)}}  + \lambda^k_1 a^{(2)}_{\ell m}&=& 0\,,\\
\frac{\partial f_k}{\partial a_{\ell m}^{(2)}}  + \lambda^k_1 a^{(1)}_{\ell m}&=& 0\,,
\end{eqnarray}
with the well known solution
\begin{equation}
f_k = -\lambda^k_1 \sum_m a_{\ell m}^{(1)} a_{\ell m}^{(2)}\,,
\end{equation}
where any value of $\lambda_1^k$ is equally good since it is simply a global factor. We have thus recovered the fact that the cross correlations are independent of the noise.

\section{Conclusions}
\label{Conclusions}

We have presented an algorithm to minimise, or even completely cancel out, the effect of systematic uncertainties (nuisance parameters) that can somehow be modelled, even if the modelling is very rough or approximate. The method was inspired by renormalization group techniques, and this interpretation provides an elegant description of what nuisance parameters are: we identify the physics that is invariant under arbitrary rescalings of the nuisance quantities. This nuisance-independent physics is completely characterized by the {\it nuisance scaling dimensions} of the observables. In the limit case where nuisance can be totally washed out this recipe will unravel the underlaying responsible {\it scale invariance}.

Our  general approach is given in Eqs. (1-2) and  we report explicit recipes for cases where the observables have  power law and linear dependences on nuisance parameters in Eqs. (11) and  (12) respectively, which apply also if only an approximate cancellation of the nuisance parameter is possible. Additional constraints on nuisance parameters or observables can be included (Eq. 14).
The algorithm is general and can be applied to any experiment or observation. However, because of our own field of expertise, we have chosen to illustrate the method with some examples drawn from astrophysics and cosmology. In doing so we have discovered some interesting new results that can be used to analyse observations from large scale galaxy and supernova surveys. For the case of baryonic acoustic oscillation experiments, we provide an optimal way to combine the observed quantities to reduce systematic uncertainties in the sound horizon scale $r_s$. For supernova surveys we have provided a method to remove a dependence of the Hubble diagram of systematic uncertainties due to metallicity variability of the SN (or, as a proxy,  that of the host galaxy). Finally, we show how for the cosmic clock method the feared age-metallicity degeneracy can be completely removed by choosing adequate combinations of the observed quantities. We chose to focus on these techniques because they are the most promising to unveil the nature of dark energy; removing or minimizing the dependence of this methods on poorly constrained systematics is crucial to be able to gain full advantage of the significant observational effort that is being invested in observational cosmology. As we have emphasized before our method is general and we hope it will be used in other areas of the experimental sciences.   

\begin{appendix}

\section{The general problem}

In this appendix we wish to suggest an interpretation of Eq.~\ref{eq:totderiv} that might prove useful when solving more general problems. As in the main text, we consider $N$ observables $O_i(\mu_1,...,\mu_m, \nu_1,...,\nu_n)$ that depend on $m$ accurate or interesting quantities $\mu_i$ and $n$ ``nuisance'' or ``biased'' parameters $\nu_i$. Let us keep the quantities $\mu_i$ fixed at some value, and assume that there are more observables than nuisance parameters $N>n$. The functions $O_i(\nu_1,...,\nu_n)$ then  define a mapping between $\mathbb{R}^n$ and $\mathbb{R}^N$, which under certain smoothness assumptions defines a manifold (an $n$-dimensional hypersurface embedded in $\mathbb{R}^N$). If this manifold is orientable, there will be at least one vector field $v^1_i$ orthogonal to it, and in general there can be up to $N - n$ such vector fields. The integrals of these vector fields (which exist at least locally) will be constant on the hypersurface
\begin{equation}
\sum_{i=1}^N v^k_i \frac{\partial O_i}{\partial \nu_j} = \sum_{i = 1}^N \frac{\partial f^k}{\partial O_i} \frac{\partial O_i}{\partial \nu_j} = 0\,,
\end{equation}
where $f_k$ is the integral of $v^k_i$. This is the same as equation \ref{eq:totderiv}. The problem is then to find such vector fields and integrate them.

\end{appendix}

\section*{acknowledgements}
It is a pleasure to thank Roland de Putter and Ben Hoyle for useful discussions. JN is supported by FP7- IDEAS Phys.LSS 240117. LV acknowledges support from FP7-PEOPLE-2007-4-3-IRG n. 202182 and FP7-  IDEAS Phys.LSS 240117. LV and  RJ are supported by MICINN grant AYA2008-03531. CPG is supported by the Spanish MICINN grant
FPA-2007-60323 and the Generalitat Valenciana grant PROMETEO/2009/116. The work of C.G. was supported in part by Grants: FPA 2009-07908, CPAN (CSD2007-00042) and HEPHACOS-S2009/ESP1473.

\end{document}